\documentclass[aps,pra,twocolumn,showpacs,floatfix]{revtex4}
\usepackage{graphicx}
\usepackage{graphics}
\usepackage{amsmath}
\begin{document}
\preprint{Version: re-submitted \today}
\title{Multimode quantum limits to the linewidth of an atom laser}
\author{Mattias T. Johnsson and Joseph J. Hope}
\affiliation{ARC Centre of Excellence for Quantum Atom Optics, The Australian National University, Canberra, 0200, Australia.}
\date{\today}

\begin{abstract}
The linewidth of an atom laser can be limited by excitation of higher energy modes in the source Bose-Einstein condensate, energy shifts in that condensate due to the atomic interactions, or phase diffusion of the lasing mode due to those interactions.  The first two are effects that can be described with a semiclassical model, and have been studied in detail for both pumped and unpumped atom lasers.  The third is a purely quantum statistical effect, and has been studied only in zero dimensional models.  We examine an unpumped atom laser in one dimension using a quantum field theory using stochastic methods based on the truncated Wigner approach.  This allows spatial and statistical effects to be examined simultaneously, and the linewidth limit for unpumped atom lasers is quantified in various limits.
\end{abstract}

\pacs{03.75.Pp, 05.10.Gg, 03.70.+k}

\maketitle

\section{Introduction}
The experimental realization of Bose-Einstein condensates (BECs) provided a testbed for many fundamental questions in interacting quantum systems, as well as providing a general tool for investigating aspects of atomic physics \cite{dalfovoET1999, ketterleET1999, leggett2001, hall2003}. One major application that BECs offer is the possibility of creating an atom laser. Atom lasers are the matter wave analogy of the optical laser, and are created by coupling atoms out of a BEC by using some external means to change the state of a subset of the atoms in the condensate from a trapped to an untrapped state \cite{mewesET1997,blochET1999,hagleyET1999}. This can produce a beam of atoms that exhibits both spatial and temporal coherence \cite{andrewsET1997, hallET1998, ottlET2005, andersonET1998, wiseman1997,sabaET2005}.
While the optical laser is well-studied and well-understood, the atom laser is still in its infancy, both in terms of a full theoretical description and experimental realization.

An important property of a laser is its linewidth, which is of crucial importance for many measurements, including spectroscopy and interferometry \cite{johnssonET2006}.  For an ``unpumped" optical laser, which is simply a cavity containing light, the linewidth is simply the inverse damping rate.  For a pumped optical laser, the linewidth is often limited by technical effects of the pumping, but the fundamental quantum limit is the Schawlow-Townes limit, which can be seen as a phase diffusion process arising from the addition of spontaneously emitted photons with random phases into the radiation field \cite{scullyET1997}. The derivation of this limit in the optical case relies on the fact that photons do not interact with each other, and consequently the damping rate of the cavity is independent of the photon population.

For an atom laser, things are more complicated. The different dispersion relations of free atoms compared to free photons means that the relationship between outcoupling and linewidth of a cavity is non-trivial, except in the weak outcoupling limit \cite{johnssonET2006}. When the Bose-Einstein condensate is stable in a single lasing mode with no interactions, the equivalent of the Schawlow-Townes limit also exists for weak outcoupling \cite{bradleyET2003}.  Unlike photons, however, atoms interact with each other strongly, resulting in a nonlinear interaction term arising from atom-atom collisions.  This has several important affects on the dynamics of an atom laser.  Semiclassical calculations show that the single-mode operation of an atom laser tend to occur in parameter regimes where the interactions are dominant \cite{haineET2002, johnssonET2005}.  Interactions also cause diffusion of the phase of the condensate, and this quantum noise is the dominant contribution to the linewidth when the single mode approximation is valid \cite{wisemanET2001}.  In practice, this phase diffusion will also be spatially dependent, and the spatial effects may interact with the effects of the quantum noise on the output.

This paper examines the linewidth of an unpumped atom laser with strong interactions, without making single mode or semiclassical approximations. The results are compared to simulations based on the semiclassical approximation and theoretical results based on a single-mode, zero-dimensional model.
We find that our results scale similarly to the semiclassical and zero-dimensional models, but the quantitative details depend on the system parameters.

General quantum field theoretic problems even in a single dimension are intractable by brute force methods, and the dominant interactions in atom laser systems means that linearized analytic calculations are not valid.  An alternative to brute force calculation is offered by stochastic phase space methods.  In this paper, we apply a truncated Wigner approach to the problem of determining the linewidth of an experimentally realistic, multimode atom laser. 

\section{Theoretical methods}
\begin{figure}[htb]
\begin{center}
\includegraphics[width=7cm,height=5cm]{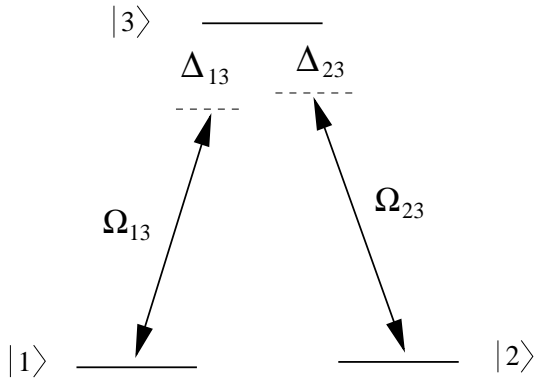}
\caption{Atomic level scheme. Using a Raman process, trapped atoms in the condensate ($|1\rangle$) are transferred to the untrapped state ($|2\rangle$) via an intermediate state ($|3\rangle$).} 
\label{fig3leveldiagram}
\end{center}
\end{figure}

Our model for the atom laser consists of a population of three level atoms in a trap, as shown in Figure \ref{fig3leveldiagram}. Bose condensed atoms in state $| 1 \rangle$ are confined by a harmonic trapping potential. These trapped atoms are then coupled to an untrapped state $|2\rangle$ in which they do not see the confining potential and consequently leave the trap forming the atom laser beam. The coupling from the trapped to the untrapped state is done via a Raman transition using the intermediate level $|3\rangle$.   Raman outcoupling has been used by two atom laser experiments \cite{hagleyET1999, robinsET2006}, and leads to an atom laser with superior properties such as a higher flux and higher brightness \cite{robinsET2006}, the ability to give the beam directionality \cite{hagleyET1999, robinsET2006}, and the possibility of creating non-classical states of the beam \cite{haineET2006}.  The Raman outcoupling scheme reduces to the rf outcoupling scheme in the limit of a zero momentum kick.
The three level system shown in Figure \ref{fig3leveldiagram} can be reduced to a two level system by adiabatically eliminating the upper level. The effective second-quantized Hamiltonian describing this reduced two-level system is given by
\begin{eqnarray}
\hat{H}_{\mathrm{eff}} &=& \int  \left[ \hat{\Psi}^{\dagger}_1 \left( -\frac{\hbar^2}{2m}\nabla^2 + \frac{1}{2} m \omega^2 r^2 - \frac{\hbar}{\Delta_{13}} |\Omega_{13}|^2 \right) \hat{\Psi}_1 \right. \nonumber \\
&& \left. + \hat{\Psi}^{\dagger}_2 \left( -\frac{\hbar^2}{2m}\nabla^2 -   \frac{\hbar}{\Delta_{13}} |\Omega_{23}|^2 - \hbar \delta \right)  \hat{\Psi}_2\right. \nonumber \\
&& + \frac{1}{2} \sum_{i=1}^2 U_{ii} \hat{\Psi}^{\dagger}_i \hat{\Psi}^{\dagger}_i \hat{\Psi}_i \hat{\Psi}_i  + U_{12}  \hat{\Psi}^{\dagger}_1 \hat{\Psi}^{\dagger}_2 \hat{\Psi}_2 \hat{\Psi}_1 \nonumber \\
&&  \left. - \hbar \left( \Omega e^{i({\mathbf{k}}_0 \cdot {\mathbf{r}} - \delta t)} \hat{\Psi}_2^{\dagger} \hat{\Psi}_1 + {\mathrm{H.c.}} \right) \right] d{\mathbf{r}}
\label{eqHeff} 
\end{eqnarray}
where $\hat{\Psi}_1({\mathbf{r}})$ and $\hat{\Psi}_2({\mathbf{r}})$ describe the trapped and untrapped matter fields respectively, $\delta = \Delta_{23} - \Delta_{13}$ is the two photon detuning, $\Omega = \Omega_{13}^* \Omega_{12}/\Delta_{13}$ is the two-photon Raman Rabi frequency, and ${\mathbf{k}}_0 = {\mathbf{k}}_2 - {\mathbf{k}}_1$ is the momentum kick imparted to the untrapped atoms from the two photon transition. The trap has been assumed to be isotropic and the trapping frequency is $\omega$. As we have included position dependence in the matter fields, the effective Hamiltonian describes the full multimode nature of the problem, and also includes non-Markovian effects.  We assume that the atomic gas is sufficiently cold and dilute, so only binary collisions are relevant and the nonlinear potentials are defined by 
\begin{equation}
U_{ij} = 4 \pi \hbar^2 a_{ij}/m,
\label{eqNonlinearPotentialDefinition}
\end{equation}
where $a_{ij}$ is the $s$-wave scattering length between atoms in state $|i\rangle$ and state $|j\rangle$.

When the quantum statistics have no effect on the dynamics of the mean field, these equations can be solved semiclassically, using the Gross-Pitaevskii (GP) equation to describe the atoms. The coupled GP equations arising from Eq.\ (\ref{eqHeff}) are given by
\begin{eqnarray}
i \hbar \frac{\partial \psi_1}{\partial t} &=& \bigg( \frac{-\hbar^2}{2m} \nabla^2 + \frac{1}{2} m \omega^2 r^2 - \frac{\hbar}{\Delta_{13}} |\Omega_{13}|^2 + U_{11}|\psi_1|^2 \nonumber \\
&&  \hspace{0.5cm} + U_{12} |\psi_2|^2 \bigg) \psi_1- \hbar \Omega e^{i {\mathbf{k}}_0 \cdot {\mathbf{r}}} \psi_2 \label{eqGPEOMpsitrapped} \nonumber\\
i \hbar \frac{\partial \psi_2}{\partial t} &=& \bigg( \frac{-\hbar^2}{2m} \nabla^2 - \frac{\hbar}{\Delta_{13}} |\Omega_{23}|^2  - \delta + U_{22}|\psi_2|^2  \nonumber \\
&& \hspace{0.5cm} + U_{12} |\psi_1|^2 \bigg) \psi_2 - \hbar \Omega^* e^{-i {\mathbf{k}}_0 \cdot {\mathbf{r}}} \psi_1  \label{eqGPEOMpsiuntrapped}
\end{eqnarray}
where we have moved to a rotating frame. These equations can be solved numerically in one, two or even three dimensions, depending on the spatial resolution required and computational resources available.  

In the absence of interactions, the field theory solution and the semiclassical solution are identical, and demonstrate that in the limit of weak interactions the output approaches a Lorentzian with a linewidth that depends on the dispersion relations of the output field \cite{johnssonET2006}.  In this Markovian limit the linewidth is the inverse of the time taken to empty the condensate.  While the proportionality constant changes depending on the potential seen by the output field, in the limit the linewidth is always proportional to the rate of the state-changing mechanism, and therefore has no lower bound.  At higher coupling rates, the timescale of the transport of the atoms from the outcoupling regime becomes comparable to the back-coupling rate, and the beam demonstrates increasingly complicated spatial and spectral behavior, up to and including the shut down of the outcoupling process altogether \cite{robinsET2005}.

In the presence of interactions, the semiclassical equations given by Eq.\ (\ref{eqGPEOMpsiuntrapped}) can be solved numerically, and demonstrate important issues even in the limit of weak outcoupling, such as a `chirp' in the frequency of the output beam due to the decay of the energy of the trapped state \cite{johnssonET2006}.  The `chirp' can be removed by adjusting the two-photon detuning during the outcoupling, but the presence of interactions also cause the quantum field to exhibit effects not seen by the semiclassical model.  The phase diffusion of the atomic field will cause a broadening of the linewidth of the atom laser, and in the limit of weak outcoupling, this will be the dominant lower bound to the linewidth.  It is also possible for these effects to have spatial effects in the high outcoupling limit, as the three timescales in the problem become comparable to each other.

We will first estimate the size of the effects of the quantum statistical effects before calculating them in a multimode model.

\subsection{Single mode limit}
\label{secSingleModeLimit}

Weakly outcoupled atom lasers use condensates that should be well described by the Thomas-Fermi limit. In this limit, the chemical potential depends on the dimension of the condensate, and is given by
\begin{eqnarray}
\mu_{1d} &=& \frac{m \omega^2}{2} \left( \frac{3 N U}{2 m \omega^2 A} \right)^{2/3} \label{mu1D} \\
\mu_{2d} &=& \left( \frac{U m \omega^2 N}{\pi l} \right)^{1/2}\label{mu2D}  \\
\mu_{3d} &=& \frac{m \omega^2}{2} \left(\frac{15 N U} {4 \pi m \omega^2} \right)^{2/5} \label{mu3D}
\end{eqnarray}
where the nonlinear potential $U$ is given by Eq.\ (\ref{eqNonlinearPotentialDefinition}), $\omega$ is the geometric mean of the trapping frequencies, and $A$ and $l$ are dimensional reduction factors corresponding to an area and a length respectively.

These dimension reduction factors arise when one has a condensate that is tightly confined in one or two dimensions, corresponding to a highly anisotropic trap.
The tight confinement along one or two axes ensures the nonlinearity has negligible effect in these directions, meaning the matter field operator will factorize with the transverse dependence completely described by a coherent state occupation of the lowest radial trap mode. This assumption leads to a second quantized Hamiltonian identical to Eq.\ (\ref{eqHeff}), except that the matter fields are one- or two-dimensional, the integral is over a length or an area rather than a volume, and the nonlinear potentials are scaled by a transverse area or transverse length. This area or length corresponds to the cross sectional area or width of the condensate along the tight trap directions \cite{steelET1998}. Dimensional reduction is a useful technique for numerical simulation, as reducing a three-dimensional problem to a one-dimensional problem greatly reduces the computation required.

If one assumes that the condensate is in a coherent state, it has one standard deviation number uncertainty given by $\sqrt{N}$. Consequently the energy uncertainty in the condensate due to number fluctuations will be given by 
\begin{equation}
\Delta E = \sqrt{N} \frac{d \mu}{d N},
\end{equation}
which means the energy uncertainty of the condensate is
\begin{eqnarray}
\Delta E_{1d} &=& \frac{m \omega^2}{3} \left( \frac{6 \pi \hbar^2 a}{m^2 \omega^2 A} \right)^{2/3} N^{1/6} \label{dE1D} \\
\Delta E_{2d} &=& \hbar \omega \sqrt{\frac{a}{l}} \label{dE2D} \\
\Delta E_{3d} &=& \frac{2 m \omega^2}{5} \left(\frac{15 \hbar^2 a} {m^2 \omega^2} \right)^{2/5} N^{-1/10}. \label{dE3D}
\end{eqnarray}
Equations (\ref{dE1D})--(\ref{dE3D}) give a rough measure of the fundamental linewidth of an atom laser, provided that the linewidth limit is due to number fluctuations in the condensate being transformed into phase fluctuations due to the nonlinear atom-atom interactions.  In order to test the accuracy of this estimate, we require a fully multimode model of the system, for which we will use stochastic methods.

\subsection{Stochastic methods}

The standard approach to using stochastic methods for a specific problem is to express the density matrix of the system in a specific basis as a quasiprobability distribution such as the $P$, positive $P$ or Wigner distribution. The master equation of the system is then converted to an equivalent partial differential equation in terms of the distribution chosen, which can be cast in form of the Fokker-Planck equation
\begin{eqnarray}
\frac{\partial P({\boldsymbol{\alpha}})}{\partial t} &=& -\sum_i \frac{\partial}{\partial \alpha_i} A_i ({\boldsymbol{\alpha}},t) P({\boldsymbol{\alpha}},t) \nonumber  \\
&& + \frac{1}{2} \sum_{ij}\frac{\partial}{\partial \alpha_i} \frac{\partial}{\partial \alpha_j} D_{ij}({\boldsymbol{\alpha}}) P({\boldsymbol{\alpha}}) 
\label{eqFokkerPlanckDefinition}
\end{eqnarray}
where ${\boldsymbol{\alpha}}$ is a vector of complex fields and $D=BB^T$ is the diffusion matrix.
Given a Fokker-Planck equation of the form (\ref{eqFokkerPlanckDefinition}), a fully equivalent formulation is given by the system of It\^o stochastic equations
\begin{equation}
\frac{\partial \alpha_i}{\partial t} = A_i + B_{ji} \eta_i(t)
\label{eqIto}
\end{equation}
where the $\eta_i(t)$ are set of noise sources with zero mean that are delta-correlated in time. It is therefore possible to describe the evolution of a quantum system in terms of a set of Langevin or stochastic partial differential equations. These stochastic equations contain all the information of the original quantum problem in that various ensemble averages of the stochastic variables correspond to various expectation values of the original quantum fields. In the case of the Wigner distribution, the stochastic average of an expression consisting of complex-valued fields corresponds to the expectation value of the symmetrized version of the quantum field operator version of that expression.

This means that if we wish to find the expectation value of some normally ordered operator $\hat{O}$ given by
\begin{equation}
\hat{O}(\hat{a},\hat{a}^{\dagger}) = \sum_{n,m} c_{nm}\hat{a}^{\dagger n} \hat{a}^m,
\end{equation}
we need only calculate the stochastic average of some c-number function $O_s(\alpha^* , \alpha)$. For the Wigner distribution, $O_s$ is given by \cite{scullyET1997}
\begin{eqnarray}
O_s(\alpha^* , \alpha) &=& \sum_{n,m} c_{nm} \left[ \frac{\partial}{ \partial (i\beta)} + \frac{\beta^*}{2i} \right]^n \nonumber \\
&& \hspace{-1cm} \times \left[ \frac{\partial}{ \partial (i\beta^*)} + \frac{\beta}{2i} \right]^m
e^{i \beta* \alpha + i \beta \alpha^*} |_{\beta^* = \beta =0}
\end{eqnarray}
and
\begin{equation}
\langle \hat{O} (\hat{a}^{\dagger}, \hat{a}) \rangle = \overline{O_s (\alpha^*, \alpha)}
\end{equation}
where the overbar indicates a stochastic average. Thus, for example, the stochastic average equivalent to the number operator is given by
\begin{equation}
\langle \hat{a}^{\dagger} \hat{a} \rangle = \overline{\alpha^* \alpha - \frac{1}{2}}.
\label{eqNumOperatorStoch}
\end{equation}

The master equation corresponding to the effective Hamiltonian (\ref{eqHeff}) is given by
\begin{equation}
\frac{d \hat{\rho}} {dt} = -\frac{i}{\hbar} \left( \hat{\rho}\, \hat{H}_{\mbox{eff}} -  \hat{H}_{\mbox{eff}} \, \hat{\rho}\right).
\label{eqMasterEquation}
\end{equation}
We determine this master equation in terms of the field operators $\hat{\Psi}_1({\mathbf{r}})$ and $\hat{\Psi}_2({\mathbf{r}})$, and make the following replacements to transform it into an equation of motion for the functional Wigner distribution:
\begin{eqnarray}
\hat{\rho} &\rightarrow & W(\psi,\psi^*) \\
\hat{\Psi} \hat{\rho} &\rightarrow & \left( \psi +\frac{1}{2} \frac{\delta}{\delta \psi^{*}} \right) W(\psi,\psi^*) \\
\hat{\rho} \hat{\Psi} &\rightarrow& \left( \psi - \frac{1}{2} \frac{\delta}{\delta \psi^{*}} \right) W(\psi,\psi^*)  \\
\hat{\Psi}^{\dagger} \hat{\rho} &\rightarrow& \left( \psi^{*} - \frac{1}{2} \frac{\delta}{\delta \psi} \right) W(\psi,\psi^*)   \\ 
\hat{\rho} \hat{\Psi}^{\dagger} &\rightarrow& \left( \psi^* + \frac{1}{2} \frac{\delta}{\delta \psi} \right) W(\psi,\psi^*). 
\end{eqnarray}
We then rearrange the result in the form of a Fokker-Planck equation in terms of the c-number stochastic variables $\psi_1$ and $\psi_2$. This gives
\begin{eqnarray}
i \hbar \frac{d W(\psi,\psi^*)}{dt} &=& \int d{\mathbf{r}} \left[ -\frac{\delta}{\delta \psi_1} \bigg( ({\cal{K}}+{\cal{V}}) \psi_1 \right.  \nonumber \\
&& \hspace{-1.5cm} - \frac{\hbar}{\Delta_{13}} |\Omega_{13}|^2 \psi_1 - U_{11} (1 - |\psi_1|^2)\psi_1 \nonumber \\
&& \hspace{-1.0cm} - U_{12} (1 - |\psi_2|^2) \psi_1
- \hbar \Omega e^{i {\mathbf{k}}_0 \cdot {\mathbf{r}}}\psi_2 \bigg) \nonumber  \\
&& \hspace{-2cm}  -\frac{\delta}{\delta \psi_2} \bigg( {\cal{K}} \psi_2 - \frac{\hbar}{\Delta_{23}} |\Omega_{23}|^2 \psi_2 - U_{22} (1 - |\psi_2|^2)\psi_2 \nonumber \\
&& \hspace{-1.5cm}  - U_{12} (1 - |\psi_1|^2) \psi_2 - \hbar \Omega^* e^{-i {\mathbf{k}}_0 \cdot {\mathbf{r}}}\psi_1 \bigg) \nonumber  \\
&& \hspace{-2cm} -\frac{\delta}{\delta \psi_1^*} \left( -({\cal{K}}+{\cal{V}}) \psi_1^* + \frac{\hbar}{\Delta_{13}} |\Omega_{13}|^2 \psi_1^* \right. \nonumber \\
&& \hspace{-1.5cm} + U_{11} (1 - |\psi_1|^2) \psi_1^* - U_{12} (1 -|\psi_2|^2)\psi_1^* \nonumber \\
&& \hspace{-1.0cm} + \hbar \Omega^* e^{-i {\mathbf{k}}_0 \cdot {\mathbf{r}}}\psi_2^* \bigg) \nonumber  \\
&& \hspace{-2cm} -\frac{\delta}{\delta \psi_2^*} \bigg( -{\cal{K}} \psi_2 + \frac{\hbar}{\Delta_{23}} |\Omega_{23}|^2 \psi^*_2 - U_{22} (1 - |\psi_2|^2)\psi_2^* \nonumber \\
&& \hspace{-1.5cm}  - U_{12} (1 - |\psi_1|^2) \psi_2^* +  \hbar \Omega e^{i {\mathbf{k}}_0 \cdot {\mathbf{r}}}\psi_1^* \bigg) \bigg]
\label{eqFokkerPlanckAtom}
\end{eqnarray}
where ${\cal{K}} = -\hbar^2 \nabla^2 /2m$, ${\cal{V}} = m \omega^2 r^2 /2$ and where the third order functional derivatives have been dropped. These third order terms do not have a simple mapping to stochastic partial differential equations, and can be assumed to be negligible when the field has a high occupation number.  This ``truncated" Wigner approach has been used successfully in a range of calculations \cite{steelET1998, drummondET2001, isellaET2005, ruostekoskiET2005, norrieET2005, norrieET2006, isellaET2006, johnssonET2006B}.

A Fokker-Plank equation can be sampled with equivalent stochastic equations where the noise terms depend only on the second order derivatives.  As Eq.\ (\ref{eqFokkerPlanckAtom}) has no second order terms,  the equivalent equations are completely deterministic once a particular initial state is chosen. Although the evolution is completely deterministic, we have not removed all effects of the quantum noise, as the noise will still enter in the choice of initial states.

Comparing Eqs.\ (\ref{eqFokkerPlanckDefinition}), (\ref{eqIto}) and (\ref{eqFokkerPlanckAtom}) we see that  Eq.\ (\ref{eqFokkerPlanckAtom}) is equivalent to the following pair of partial differential equations
\begin{eqnarray}
i \hbar \frac{\partial \psi_1}{\partial t} &=& \left( \frac{-\hbar^2}{2m} \nabla^2 + \frac{1}{2} m \omega^2 r^2 - \frac {\hbar |\Omega_{13}|^2} {\Delta_{13}} \right. \nonumber \\
&&  + U_{11}(|\psi_1|^2 - \frac{1}{\Delta V})  + U_{12} (|\psi_2|^2 - \frac{1}{2\Delta V}) \Big) \psi_1 \nonumber \\
&& - \hbar \Omega e^{i {\mathbf{k}}_0 \cdot {\mathbf{r}} } \psi_2  \label{eqStochPDEpsitrapped} \\
i \hbar \frac{\partial \psi_2}{\partial t} &=& \left( \frac{-\hbar^2}{2m} \nabla^2 - \frac{\hbar |\Omega_{23}|^2} {\Delta_{13}}   - \hbar \delta + U_{22}(|\psi_2|^2 -\frac{1}{\Delta V}) \right. \nonumber \\
&& \left. + U_{12} (|\psi_1|^2 -\frac{1}{2\Delta V}) \right) \psi_2 - \hbar \Omega^* e^{-i {\mathbf{k}}_0 \cdot {\mathbf{r}}} \psi_1 \label{eqStochPDEpsiuntrapped}
\end{eqnarray}
where $\Delta V$ is the volume element of the discretization of the problem.  The terms inversely proportional to these volume elements compensate for the mean field of the vacuum, which is non-zero in the Wigner approach.  Apart from these terms, these equations are identical to the semiclassical equations for the system given in Eq.~(\ref{eqGPEOMpsiuntrapped}).

To determine the noise distribution we need to apply to our initial states, we make the assumption that the condensate is in a multimode coherent state, which is equivalent to assuming that each of the single mode fields at each of the grid points $x_i$ is in a single mode coherent state $|\psi^0_j(x_i)\rangle$. We can find an appropriate ground coherent state for the trapped atoms by integrating the semiclassical equations in imaginary time, and the output is assumed to be in the vacuum state initially.

The Wigner distribution for a single mode coherent state $|\alpha \rangle$ is given by 
\begin{equation}
W(y_1, y_2) = \frac{2}{\pi} \exp \left[ -2 \left( (x_1 - \alpha_r)^2 + (x_2 - \alpha_i)^2 \right) \right]
\label{eqWignerNotSqueezed}
\end{equation}
where $\alpha_r$ and $\alpha_i$ are the real and imaginary parts of the coherent amplitude $\alpha$, and $y_1$ and $y_2$ correspond to the real and imaginary parts of the atomic field. This Wigner function represents a Gaussian uncertainty with a standard deviation of one half along both the $y_1$ and $y_2$ axes. 
Thus, to give our initial fields the correct statistics, at each grid point we need to apply noise of the following form:
\begin{eqnarray}
\psi_1(x_i) &=& \psi^{0}_1(x_i) + \frac{\eta_1(x_i)}{\sqrt{\Delta V}} \\
\psi_2(x_i) &=& \psi^{0}_2(x_i) + \frac{\eta_2(x_i)}{\sqrt{\Delta V}}
\end{eqnarray}
where the $\eta_j(x_i)$ are complex Gaussian noise functions with a standard deviation in the real and imaginary components of one half.  

To investigate the linewidth behavior, we solved the stochastic equations (\ref{eqStochPDEpsitrapped}) and (\ref{eqStochPDEpsiuntrapped}) numerically, using the mathematical package XMDS \cite{XMDS}. This open source package allows the solution of systems of both stochastic and deterministic partial differential equations on grids of arbitrary dimension. It also allows the use of MPI methods to distribute the computation over many processors, an approach that is crucial when carrying out multi-trajectory simulations such as those required by stochastic PDEs. These stochastic solutions were then compared with the semiclassical solutions obtained by solving Eq.~(\ref{eqGPEOMpsiuntrapped}).

The linewidth of the the atom laser beam is the width of the momentum distribution of the beam as a function of time. Using Eq.~(\ref{eqNumOperatorStoch}) along with the linearity of the Fourier transform, one can show that the expression for the beam momentum in terms of the stochastic variables is given by   
\begin{equation}
\langle \hat{\Psi}^{\dagger}_2(k) \hat{\Psi}_2(k) \rangle = \overline{|\psi_2 (k)|^2 - \frac{1}{2 \Delta V_k}}
\label{eqMomentumSpacePowerDensity}
\end{equation}
where $\Delta V_k$ is the volume element of the discretized grid in momentum space.

The system was mostly solved in one dimension for reasons of computational efficiency, although a limited number of two-dimensional simulations were carried out in order to ensure that the scaling laws carried over from one dimension to two as expected.  The dimensional reduction of the problem from three dimensions to one was performed as described in Section \ref{secSingleModeLimit}. We chose a transverse area of $A=1.2\times 10^{-11}\,$m$^2$ for reduction to a 1D problem and a transverse length of $l=3.46 \times 10 ^{-6}\,$m for the reduction to a 2D problem.

We focus on the weak outcoupling regime in these simulations so that we can examine the effects of interaction-induced phase diffusion in isolation from other limits to the linewidth.  In the strong outcoupling regime a number of effects can cause density fluctuations in the condensate which are reproduced in the output beam \cite{robinsET2005}, and this classical noise dominates the resulting linewidth. Also, the chemical potential decreases with the number of atoms in the condensate, resulting in a smaller mean field kick to the atoms as they leave the trap. This causes a ``chirp'' in the momentum distribution of the beam, artificially broadening the linewidth \cite{johnssonET2006}.  In the weak outcoupling limit, the interaction-induced phase diffusion is the dominant limit to the linewidth.

\section{Results}

\begin{figure}[htb]
\begin{center}
\includegraphics[width=4.2cm,height=4.2cm]{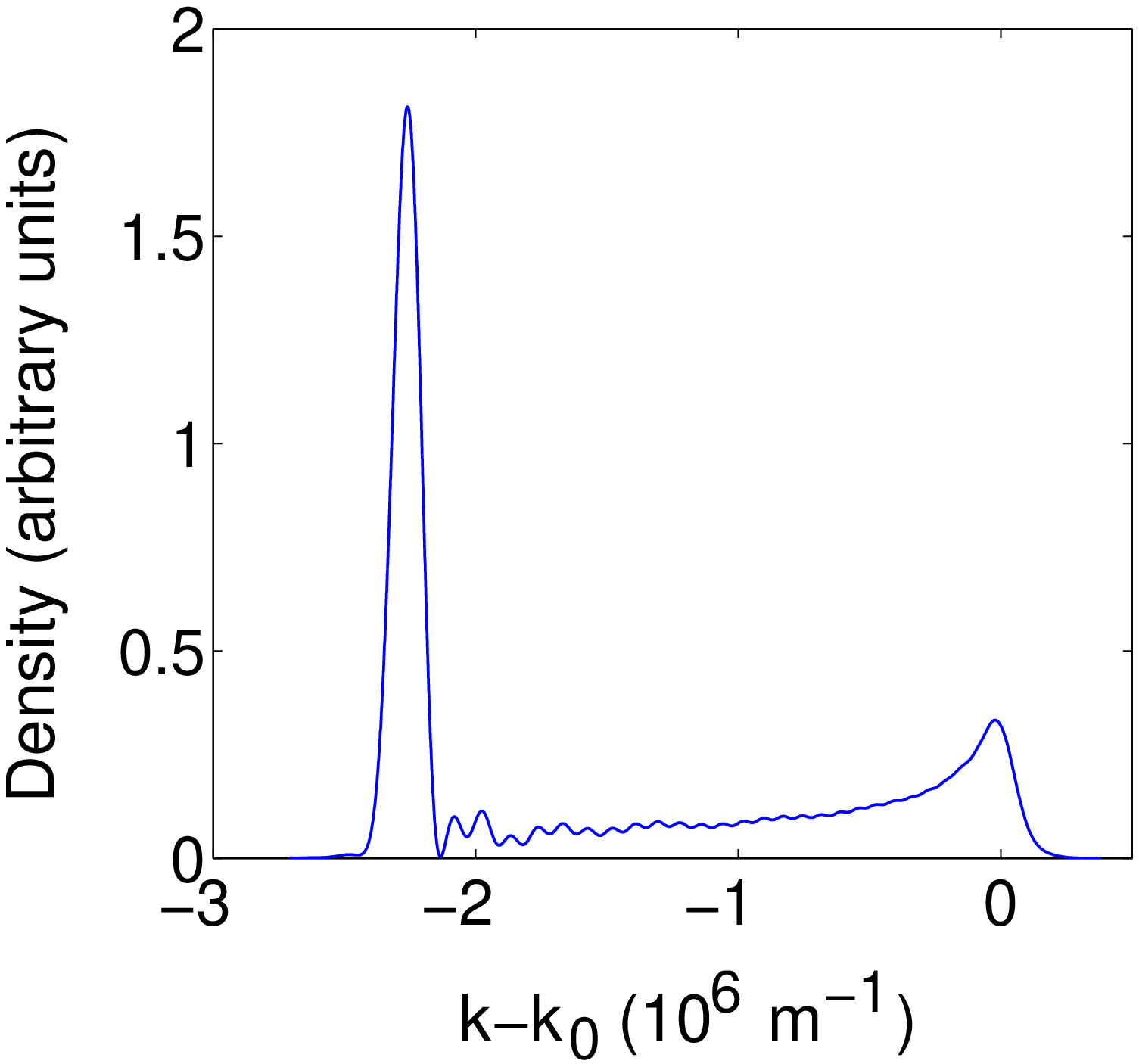}
\includegraphics[width=4.2cm,height=4.2cm]{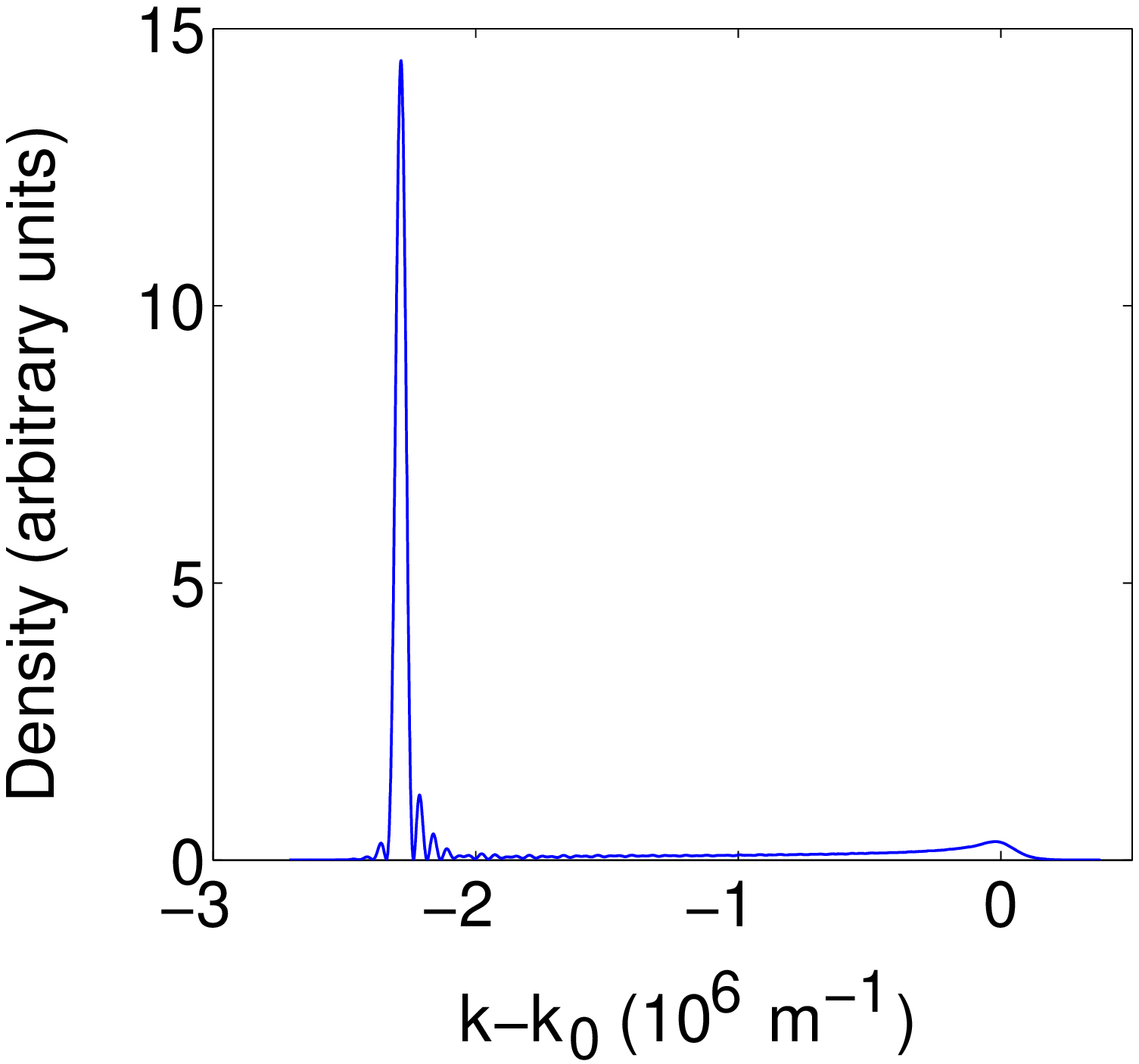}
\caption{Momentum space power spectrum of the output beam after 5ms (left) and 10ms (right). Over time, the atoms build up outside the coupling region, and gain a small extra momentum kick due to the BEC-beam repulsion.  Parameters: $N=5\times 10^{6}$, $\omega = 250\,$rad$\,$s$^{-1}$, $a=1\times 10^{-9}\,$m, $k_0 = 2\times 10^{7}\,$m$^{-1}$.} 
\label{figkSpaceSpectrum5msAnd10ms}
\end{center}
\end{figure}

The quantum noise has no effect on the short term behavior of the atom laser.  The main short-term features are shown in Figure \ref{figkSpaceSpectrum5msAnd10ms}, which shows the outcoupled field in momentum space.  In the first few microseconds, the output momentum spectrum mimics the trapped momentum spectrum shifted to be centered around the total momentum kick from the Raman transition $\hbar k_0$.  As the atoms are accelerated out of the condensate due to the mean field repulsion, this peak moves, and over time the energy width of the output field reduces and this peak narrows.  Thus, just after outcoupling begins, there is a spread of momenta in the beam ranging from the Raman kick $\hbar k_0$ up to the Raman kick plus the momentum corresponding to the energy $\mu$ gained from the mean field.  As time progresses, many more atoms accumulate in the part of the beam that is outside the condensate rather than in the outcoupling region within the condensate.  Consequently, in momentum space, most of the momentum is now concentrated in a peak at $\hbar k  = \sqrt{\hbar^2 k_0^2 + 2m \mu}$, and it is the width of this peak that we will consider to be the linewidth of the atom laser.  Early snapshots of this behavior are shown in Figure \ref{figkSpaceSpectrum5msAnd10ms}. 

Previous semiclassical analyses of the width of this peak have shown that it is essentially Fourier limited: that is, its width is inversely proportional to the outcoupling time \cite{johnssonET2006}. This narrowing continues indefinitely according to a semiclassical analysis, a result that must inevitably fail at some point due to quantum statistical fluctuations of the phase of the BEC providing the atom source.  The quantum statistics become important on a timescale depending on the strength of the interactions, which was typically on the order of tens of milliseconds in our simulations. 

\begin{figure}[htb]
\begin{center}
\includegraphics[width=7cm,height=6cm]{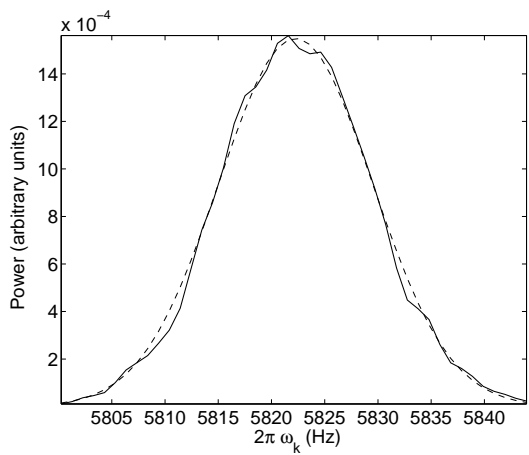}
\caption{Power spectrum of the output beam at $t=0.35\,s$ according to the stochastic simulation (solid line) as well as a Gaussian fitting profile (dashed line). Parameters: $N=10^{7}$, $\omega = 250\,$rad$\,$s$^{-1}$, $a=3\times 10^{-9}\,$m, $k_0 = 10^{7}\,$m$^{-1}$.} 
\label{figFrequencySpectrumStochasticGaussianFit0.35s}
\end{center}
\end{figure}

\begin{figure}[htb]
\begin{center}
\includegraphics[width=7cm,height=6cm]{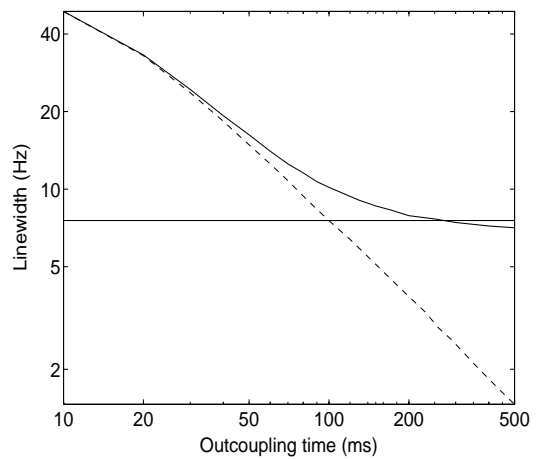}
\caption{Linewidth narrowing as a function of outcoupling time. Dashed line shows the semiclassical result; solid line shows the result of the stochastic simulation. Horizontal line indicates the fundamental linewidth limit according to twice Eq.\ (\ref{dE1D}). Parameters: $N=10^{7}$, $\omega = 250\,$rad$\,$s$^{-1}$, $a=1\times 10^{-9}\,$m, $k_0 = 10^{7}\,$m$^{-1}$.} 
\label{figLinewidthNarrowing_a1nm_loglog}
\end{center}
\end{figure}

\begin{figure}[htb]
\begin{center}
\includegraphics[width=7cm,height=6cm]{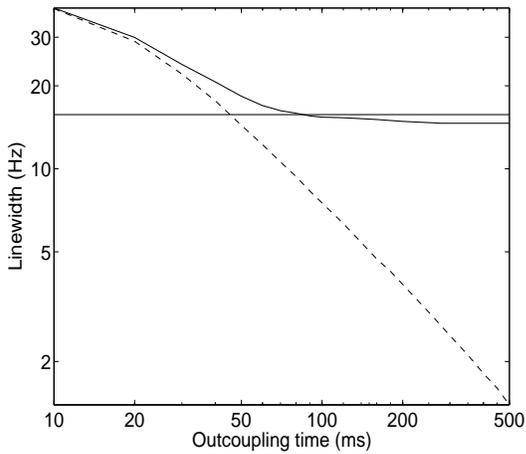}
\caption{Linewidth narrowing as a function of outcoupling time. Dashed line shows the semiclassical result; solid line shows the result of the stochastic simulation. Horizontal line indicates the fundamental linewidth limit according to twice Eq.\ (\ref{dE1D}). Parameters: $N=10^{7}$, $\omega = 250\,$rad$\,$s$^{-1}$, $a=3\times 10^{-9}\,$m, $k_0 = 10^{7}\,$m$^{-1}$.} 
\label{figLinewidthNarrowing_a3nm_loglog}
\end{center}
\end{figure}

The momentum space power spectrum of the output beam was calculated by using Eq.\ (\ref{eqMomentumSpacePowerDensity}) and taking the stochastic average over 1024 paths unless otherwise stated. The resulting power spectrum of the beam was then fit to a Gaussian envelope function and the full width, $1/e$ height taken as the linewidth. This definition of linewidth corresponds to twice the standard deviation of the Gaussian, and as such is given by twice the energy uncertainty given in Eqs.\ (\ref{dE1D})--(\ref{dE3D}).  Due to the finite number of paths and the stochastic nature of the problem, the Gaussian fit is not exact, but it still resembles a Gaussian very closely. An example is shown in Figure \ref{figFrequencySpectrumStochasticGaussianFit0.35s}.
When we examine the narrowing of the laser linewidth using both a semiclassical and a stochastic approach, we notice a striking difference in long-term behavior. The linewidth of the beam over time is plotted for two different parameter regimes in Figure \ref{figLinewidthNarrowing_a1nm_loglog} and Figure \ref{figLinewidthNarrowing_a3nm_loglog}. These Figures show that both the semiclassical and stochastic results agree initially on the rate at which the linewidth narrows, but at later times the stochastic simulation shows that the linewidth hits a limit, while in the semiclassical simulation it continues to narrow. 

In the long time limit the shape of the semiclassical (dashed) curve in Figures  \ref{figLinewidthNarrowing_a1nm_loglog} and \ref{figLinewidthNarrowing_a3nm_loglog} is linear with a slope of -1, indicating that linewidth is inversely proportional to the outcoupling time. This agrees with the Fourier arguments in \cite{johnssonET2006}.  The fundamental linewidth limit due to interaction-induced phase diffusion for an approximate single mode model, given by twice Eq.~(\ref{dE1D}), is shown on the figures as a horizontal bar, and agrees closely with the results of the multimode stochastic simulations.

\begin{figure}[htb]
\begin{center}
\includegraphics[width=7cm,height=6cm]{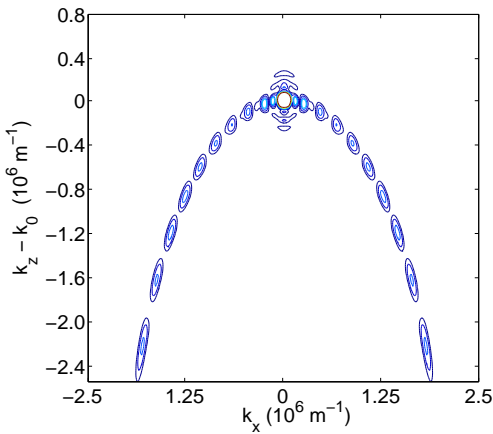}
\caption{Momentum space density of a two dimensional atom laser, with $p_x = \hbar k_x$, $p_z = \hbar k_z$. The parabolic shape is due to energy conservation as per Eq.\ (\ref{eqMomentumSpaceArc}), with the thickness of the line tracing out the parabola giving the linewidth. Parameters: $N=5 \times 10^{6}$, $\omega = 1500\,$rad$\,$s$^{-1}$, $a=1\times 10^{-8}\,$m, $k_0 = 4 \times 10^{6}\,$m$^{-1}$.} 
\label{fig2DMomentumSpacePlot}
\end{center}
\end{figure}

\begin{figure}[htb]
\begin{center}
\includegraphics[width=7cm,height=6cm]{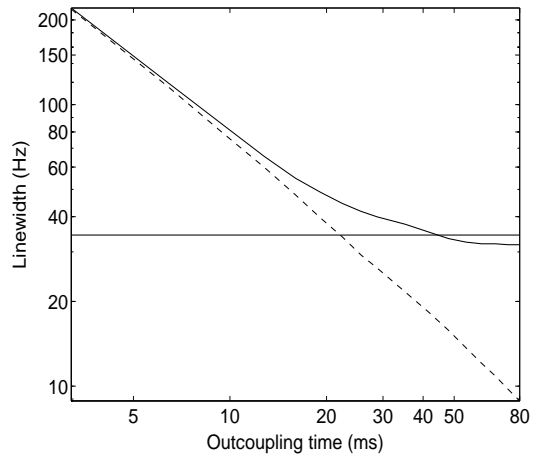}
\caption{Results for a two dimensional simulation. Linewidth narrowing as a function of outcoupling time, with energy spread measured in the longitudinal direction. Dashed line shows the semiclassical result; solid line shows the result of the stochastic simulation. Horizontal line indicates the fundamental linewidth limit according to (\ref{dE2D}). Parameters: $N=5 \times 10^{6}$, $\omega = 2000\,$rad$\,$s$^{-1}$, $a=1\times 10^{-8}\,$m, $k_0 = 2 \times 10^{7}\,$m$^{-1}$.} 
\label{figLinewidthNarrowing2D_a10nm_loglog}
\end{center}
\end{figure}

To check the possibility that the good agreement between the linewidth limit given by (\ref{dE1D}) and the results of the stochastic simulation is related to the restricted dimensionality of the one dimensional simulations, we repeated the simulation in two dimensions using 256 trajectories. In two dimensions the momentum distribution is now no longer a simple peak, but rather an arc in momentum space satisfying the energy conservation relation
\begin{equation}
p_x^2 + p_z^2 = \hbar^2 k_0^2 + 2 m \mu
\label{eqMomentumSpaceArc}
\end{equation}
where $x$ is the transverse degree of freedom that we have added. An example of this two-dimensional momentum space density is shown in Figure \ref{fig2DMomentumSpacePlot}. The peaks in the momentum distribution are due to transverse structure arising from interference effects in the beam and the outcoupling process.  
The parabolic arc has a specific thickness that narrows, and corresponds to the energy linewidth of the output beam. The results of this narrowing as a function of time are shown in Figure \ref{figLinewidthNarrowing2D_a10nm_loglog}. Again, the fundamental lower linewidth limit closely tracks that predicted by (\ref{dE2D}).  
%\begin{figure}[htb]
%\begin{center}
%\includegraphics[width=8cm,height=7cm]{2DMomentumSpacePlot.eps}
%\caption{2DMomentumSpacePlot} 
%\label{test1}
%\end{center}
%\end{figure}
%\begin{figure}[htb]
%\begin{center}
%\includegraphics[width=9cm,height=7cm]{2DMomentumSpacePlot2.eps}
%\caption{2DMomentumSpacePlot} 
%\label{test2}
%\end{center}
%\end{figure}
Consequently the argument that number fluctuations coupling to energy fluctuations cause a fundamental limit to the linewidth appears to scale correctly across dimensions, indicating that Eqs.\ (\ref{dE1D})--(\ref{dE3D}) will correctly estimate the fundamental linewidth of a real atom laser experiment in three dimensions. 

\section{Number squeezing}

As the linewidth of the atom laser scales with the size of the number uncertainty of the condensate, it is natural to consider to what extent the linewidth could be reduced by minimizing number uncertainty via number squeezing.

The most straightforward approach is to consider quadrature squeezing. In the single-mode case one defines amplitude and phase quadrature operators by
\begin{eqnarray}
\hat{X}^+ &=& e^{i\phi} \hat{a} + e^{-i\phi} \hat{a}^{\dagger} \\
\hat{X}^- &=& i \left( e^{i\phi} \hat{a} - e^{-i\phi} \hat{a}^{\dagger} \right) 
\end{eqnarray}
where $\phi$ is the phase angle at which the measurement is carried out. The variances of $\hat{X}^{\pm}$ are unity for a coherent state, and consequently a state is squeezed if the variance of one the quadrature operators is less than one. Squeezing effectively repartitions the unavoidable uncertainties associated with simultaneous measurement of a pair of non-commuting observables, increasing the uncertainty in one of the observables in order to reduce uncertainty in the other. In the case of quadrature squeezing, the two observables are amplitude and phase.  Using typical notation, the squeezing in a state can be characterised by a parameter $r$, where the variances of the two quadratures are given by
\begin{eqnarray}
\mbox{var} (\hat{X}^+) &=& e^{-2r} \label{eqXplusSqueezed} \\
\mbox{var} (\hat{X}^-) &=& e^{2r}. \label{eqXminusSqueezed}
\end{eqnarray}
For our purposes, the relevant property of this state is that it exhibits number squeezing.  The expectation value and variance for the number operator $\hat{N} = \hat{a}^{\dagger} \hat{a}$ are given by
\begin{eqnarray}
\langle \hat{N} \rangle &=& |\alpha|^2 + \sinh ^2 r, \label{eqNumberExpectationSqueezed} \\
\mbox{var} (\hat{N}) &=& |\alpha \cosh r - \alpha^* e^{-i \theta} \sinh r |^2 \nonumber \\
&& \hspace{1.5cm} + 2 \cosh ^2 r  \sinh ^2 r \label{eqNumberVarianceSqueezed}
\end{eqnarray}
where $\alpha$ is the coherent amplitude of the squeezed state.

In order to stochastically simulate an atom laser sourced from a quadrature squeezed BEC, we require initial noise that represents such a squeezed state, albeit the multimode rather than single mode version. Analogous to Eq.\ (\ref{eqWignerNotSqueezed}), the Wigner distribution for a squeezed coherent state is given by 
\begin{eqnarray}
W(x_1, x_2) &=& \frac{2}{\pi} \exp \left[ -2 \left( (x_1 - \alpha_r)^2 e^{-2r} \right. \right. \nonumber \\
&& \hspace{2cm} \left. \left. + (x_2 - \alpha_i)^2 e^{2r} \right) \right].
\end{eqnarray}
This noise is applied to the initial fields at each of our grid points as before.
For our simulations we chose a squeezing parameter $r=\ln 2$, which results in a number variance of $\mbox{var} (\hat{N}) = N/4$. This reduction in the number variance by a factor of four over that of a coherent state corresponds to reducing the standard deviation, and hence the linewidth, by a factor of two.

\begin{figure}[htb]
\begin{center}
\includegraphics[width=7cm,height=6cm]{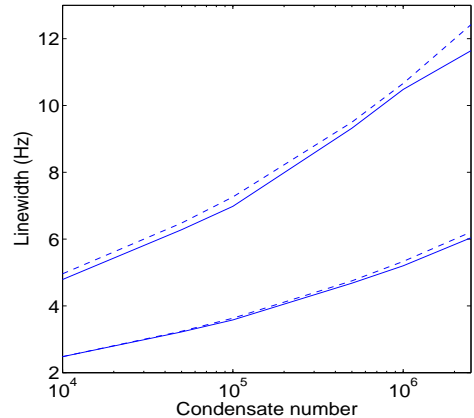}
\caption{Lower bound on linewidth as a function of condensate number for squeezed and non-squeezed condensates. Solid lines represent stochastic simulations; dashed lines the prediction according to Eqs.\ (\ref{dE1D}) and (\ref{eqNumberVarianceSqueezed}). The upper pair of lines assume a non-squeezed condensate; the lower pair represent a condensate with quadrature squeezing of magnitude $r=0.69$.} 
\label{figLinewidthNDependence}
\end{center}
\end{figure}

Figure \ref{figLinewidthNDependence} shows the linewidth of the atom laser in the long time limit over a range of BEC atom numbers. The results of stochastic simulations calculating the linewidth for a laser sourced from both squeezed and non-squeezed condensates are shown, along with the theoretical prediction for the linewidth based on Eqs.\ (\ref{dE1D}) and (\ref{eqNumberVarianceSqueezed}). Figure \ref{figLinewidthNDependence} demonstrates that the linewidth limit scales as $N^{1/6}$ as predicted, and that the reduction in number uncertainty due to number squeezing reduces the linewidth limit in agreement with Eq.\ (\ref{eqNumberVarianceSqueezed}).

\section{Conclusions}

We have quantified the lower bound for the linewidth of an atom laser due to the quantum diffusion of the BEC mode.  We have simulated this process in one and two dimensions, showing that the resulting linewidth limit is slightly lower than that estimated from a single mode BEC undergoing phase diffusion, but that the result scales in the same way as the zero-dimensional model.

Just as the semiclassical model exhibits complicated spatial behavior when the physical timescales of the outcoupling and physical transport of the atoms from the outcoupling regime become comparable, so we might expect significant spatial effects  due to the quantum noise in the system as the coupling rate is increased.  This rich behavior due to the interaction of the three timescales in the problem is essentially undesirable when considering an atom laser as an atom source, effectively resulting in an unpredictable atomic flux and energy spectrum.  It will therefore be desirable to find ways to minimize this linewidth limit while remaining in the low outcoupling rate limit.  This is a complicated trade-off: pumped systems tend to be more stable and narrow when operating at high flux, and high flux implies either high atom densities (and therefore interactions), or high outcoupling rate, where the linewidth limit is higher.

One obvious solution might be to attempt to reduce atomic interactions via some process such as a Feshbach resonance \cite{robertsET1998}, but this introduces a further complication, which is that earlier work has suggested that pumping schemes will tend to lead to modal instabilities except in the regimes of high interactions \cite{haineET2002,johnssonET2005}.  Further studies are required in order to determine the optimum way of finding a compromise between these competing processes.  By the very nature of this compromise, these studies will require a quantum statistical model of the atom laser system with pumping that includes at least one spatial dimension.  The model and results in this paper are an important step in that direction.

\section{Acknowledgments}
We would like to thank Simon Haine and Nick Robins for helpful discussions and suggestions. This work was supported by the Centre of Excellence program of the Australian Research Council and the APAC National Supercomputing Facility.

\end{document}